\documentclass[epj]{svjour}
\usepackage{graphics}
\usepackage{amsmath}
\usepackage[dvips]{graphicx}
\usepackage{epsfig}
\newcommand{\car}{\mbox{$^{12}{\mathrm C}$}}
\newcommand{\mg}{\mbox{$^{24}{\rm Mg}$}}

\newcommand{\rmax}{\mbox{$R_{\rm max}$}}
\newcommand{\ane}{\mbox{$\alpha+^{20}$Ne}}
\newcommand{\oaa}{\mbox{$^{16}{\rm O}+\alpha+\alpha$}}

\begin{document}
\title{Resonances in $\car$ and $\mg$: what do we learn from a microscopic cluster theory?}
\author{P. Descouvemont
\thanks{\emph{Directeur de Recherches FNRS}}%
}                     
\institute{Physique Nucl\'eaire Th\'eorique et Physique Math\'ematique, C.P. 229, Universit\'e Libre de Bruxelles (ULB), B 1050 Brussels, Belgium}
\date{Received: date / Revised version: date}
%
\abstract{
We discuss resonance properties in three-body systems, with examples on $\car$ and $\mg$. We use a
microscopic cluster model, where the generator coordinate is defined in the hyperspherical formalism. 
The $\car$ nucleus is described by an
$\alpha+\alpha+\alpha$ structure, whereas $\mg$ is considered as an $\oaa$ system. We essentially pay
attention to resonances. We review various techniques which may extend variational methods to resonances.
We consider $0^+$ and $2^+$ states in $\car$ and $\mg$. We show that the r.m.s. radius of a resonance is
strongly sensitive to the variational basis. This has consequences for the Hoyle state ($0^+_2$ state in $\car$) whose
radius has been calculated or measured in several works. In $\mg$, we identify two $0^+$ resonances slightly
below the three-body threshold.
%
} 
\maketitle
\section{Introduction}
\label{intro}
Clustering is a well established phenomenon in light nuclei (see Refs.\ \cite{FHK18,HIK12,DD12} for recent reviews).
In nuclear physics, most cluster states involve the $\alpha$ particle. Due to its large binding
energy, the $\alpha$ particle tends to keep its own identity, which leads to the $\alpha$ cluster structure \cite{Br66,FHI80}.
The cluster structure in $\alpha$ nuclei (i.e. with nucleon numbers $A=4k$, where $k$ is an integer number) was clarified by Ikeda
{\sl et al.} \cite{ITH68} who proposed a diagram which identifies situations where a cluster 
structure can be observed. The $\alpha$ cluster structure is essentially found
for nuclei near $N=Z$.
The $\alpha$ model and its extensions were utilized by many authors to investigate the properties of $\alpha$-particle nuclei such as $^8$Be, $^{12}$C, $^{16}$O, etc. 
In particular, the interest for $\alpha$-cluster models was recently revived by  the hypothesis of a new form 
of nuclear matter, in analogy with the Bose-Einstein condensates \cite{THS01,ZFH20}.

Nuclear models are essentially divided in two categories: (1) in non-microscopic models, the
internal structure of the clusters is neglected, and they interact by a nucleus-nucleus potential;
(2) in microscopic models, the wave functions depend on the $A$ nucleons of the system, and
the Hamiltonian involves a nucleon-nucleon interaction. Recent developments in nuclear models \cite{NF08,NQS09,He20}
aim to find exact solutions
of the $A$-body problem, but they present strong difficulties when the nucleon number increases. To simplify the problem, cluster
models assume that the nucleons are grouped in clusters. The simplest variant is a two-cluster
model and is being developed since more than 40 years \cite{WT77,Ho77}. Multicluster
microscopic models are more recent (see, for example, Ref.\ \cite{DD12}). They allow to extend
the range of applications.

In the present paper, we focus on two $\alpha$ nuclei: $\car$ and $\mg$, which are described by
three-body structures ($3\alpha$ and $\oaa$, respectively). Over the last 20 years, there was a strong
interest on $\car$, and in particular on the $0^+_2$ resonance, known as the Hoyle state \cite{FF14}.
The unbound nature, however, make theoretical studies delicate.

With $\car$ and $\mg$ as typical examples, we discuss more specifically the determination of
resonance properties in cluster models. As resonances are unbound, a rigorous treatment would
require a scattering model, with scattering boundary conditions. There are, however, various
techniques aimed at complementing the much simpler variational method which is, strictly speaking, valid
for bound states only. In a variational method, negative energies are associated with physical
bound states. The positive eigenvalues correspond to approximations of the continuum. For narrow
resonances, a single eigenvalue is in general a fair approximation. We show, however, that the 
calculation of physical quantities, such as the r.m.s. radius, should be treated carefully.

The paper is organized as follows. In Sec.\ \ref{sec2}, we present the microscopic three-body model,
using hyperspherical coordinates. Section \ref{sec3} is devoted to a brief discussion of different
techniques dealing with resonances. The $\car$ and $\mg$ nuclei are presented in Sects.\ \ref{sec4}
and \ref{sec5}, respectively. Concluding remarks are given in Sect.\ \ref{sec6}.

\section{The Microscopic three-cluster model}
\label{sec2}
\subsection{Hamiltonian and wave functions}
In a microscopic model, the Hamiltonian of a $A$-nucleon system is given by 
\begin{align}
H=\sum_{i=1}^A T_i +\sum_{i<j=1}^A(V^N_{ij}+V^C_{ij}),
\label{eq1}
\end{align}
where $T_i$ is the kinetic energy of nucleon $i$, and $V^N_{ij}$ and $V^C_{ij}$ are the nuclear and Coulomb interactions between nucleons $i$ and $j$.  

In a partial wave $J\pi$, the wave function $\Psi^{JM\pi}$ is a solution of the Schr\"odinger equation
\begin{align}
H\Psi^{JM\pi}=E\Psi^{JM\pi}.
\label{eq2}
\end{align}
Recent {\sl ab initio} models \cite{NF08,NQS09}  aim at determining exact solutions of Eq.\ (\ref{eq2}). For instance, the No-Core Shell Model (NCSM) is based on 
large one-center harmo\-nic-oscil\-lator (HO) bases and  effective interactions \cite{NKB00},
derived from realistic forces such as Argonne \cite{WSS95} or  CD-Bonn \cite{Ma01}. These interactions
are adapted for finite model spaces through a particular unitary transformation.
Wave functions are then expected to be accurate, but states presenting a strong 
clustering remain difficult to describe with this approach. 

In cluster models, the nucleon-nucleon interaction $V^N_{ij}$ must account 
for the cluster approximation of the wave function. This leads to effective interactions, 
adapted to harmonic-oscillator orbitals. For example, using $0s$ orbitals for the $\alpha$ particle 
makes all matrix elements of non-central forces equal to zero. The effect of non-central components 
is simulated by an appropriate choice of the central effective interaction. 
Typical effective interactions are the Minnesota \cite{TLT77} or the Volkov potentials \cite{Vo65}. These central forces
simulate the effects of the missing tensor interaction. They include an adjustable parameter, which can be slightly modified
without changing the basic properties of the force. This parameter is typically used to reproduce
the energy of the ground state or of a resonance.

In the present model, Eq.\ (\ref{eq2}) is solved by using the cluster approximation, i.e. the nucleus is seen as a three-body system, where each cluster is represented by a shell-model wave function.  This leads to the Resonating Group Method (RGM, see Refs.\ \cite{Ho77,DD12}) which was initially developed for two-cluster systems, but more recently extended to three-body nuclei
\cite{KD04}.  

\begin{figure}[htb]
	\begin{center}
		\epsfig{file=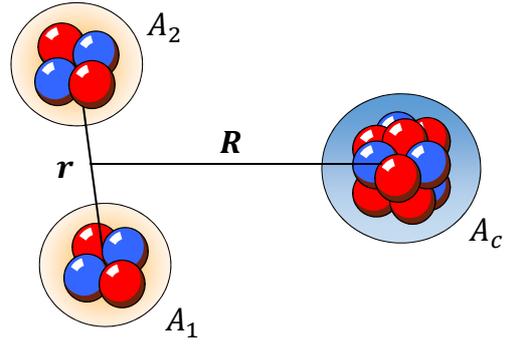,width=6.5cm}
		\caption{Coordinates in the microscopic three-cluster system.}
		\label{fig_config}
	\end{center}
\end{figure}

The three-body system is illustrated in Fig.\ \ref{fig_config}. Coordinate $\pmb{r}$ is associated with the
external clusters, whereas $\pmb{R}$ is the relative coordinate between the core and the two-body subsystem.
Scaled Jacobi coordinates $\pmb{x}$ and $\pmb{y}$ \cite{ZDF93,KD04} are obtained from
\begin{align}
	&\pmb{x}=\sqrt{A_1A_2/(A_1+A_2)}\, \pmb{r}, \nonumber \\
	&\pmb{y}=\sqrt{A_c(A_1+A_2)/(A_c+A_1+A_2)}\, \pmb{R},
\end{align}
where $A_c$ is the mass of the core and ($A_1,A_2$) the masses of the external clusters.
We use the hyperspherical formalism where the hyperradius $\rho$ and the hyperangle $\alpha$ are defined as
\begin{align}
&\rho^2=\vec{x}^2+\vec{y}^2 \nonumber \\
&\alpha=\arctan(y/x).
\label{eq3}
\end{align}
The hyperspherical formalism is well known in non-mi\-cros\-copic three-body systems \cite{KRV08,ZDF93},
where the structure of the nuclei is neglected. This formalism makes use of five angles $\Omega^5=(\Omega_x,\Omega_y,\alpha)$ 
and of an hypermomentum $K$ which generalizes 
the concept of angular momentum in two-body systems. The hyperspherical functions are \cite{RR70}
\begin{align}
{\cal Y}^{\ell_x \ell_y}_{KLM}(\Omega^{5})=\phi_K^{\ell_x \ell_y}(\alpha)
\left[ Y_{\ell_x}(\Omega_{x})\otimes Y_{\ell_y}(\Omega_{y}) \right] ^{LM},
\label{eq4}
\end{align}
where $\ell_x$ and $\ell_y$ are angular momenta associated with the Jacobi coordinates $\pmb{x}$ and $\pmb{y}$. 
Functions $\phi_K^{\ell_x \ell_y}$ are given by
\begin{align}
\phi_K^{\ell_x \ell_y}(\alpha)={\cal N}_K^{\ell_x \ell_y} \,(\cos \alpha)^{\ell_x} (\sin \alpha)^{\ell_y}
P_n^{\ell_y+\frac{1}{2},\ell_x+\frac{1}{2}}(\cos 2\alpha).
\label{eq5}
\end{align}
In these definition, $P_n^{\alpha, \beta}(x)$ is a Jacobi polynomial, 
${\cal N}_K^{l_x l_y}$ is a normalization factor, and $n=(K-\ell_x-\ell_y)/2$ is a positive integer. 

For the sake of clarity, we limit the presentation to clusters with spin 0 (a generalization can be found in Ref.\ \cite{De19}). The total wave function of the system is written as
\begin{align}
	\Psi^{JM\pi}=
	\sum_{\ell_x \ell_y}\sum_{K=0}^{\infty}  {\cal A} \,\phi_1 \phi_2 \phi_3
  {\cal Y}^{\ell_x \ell_y}_{KJM}(\Omega^{5}) \chi^{J\pi}_{\ell_x \ell_y K}(\rho),
	\label{eq6}
\end{align}
where  ${\cal A}$ is the $A$-nucleon antisymmetrizor, and
$\phi_i$ are the cluster wave functions, defined in the shell-model. 
For the $\alpha$ particle, the internal wave function $\phi$ is a Slater determinant involving
four $0s$ orbitals. Definition (\ref{eq6}), however, is valid in a broader context where $\phi$ is a linear
combination of several Salter determinants. A recent example \cite{De19} is the $^{11}$Li nucleus described
by a $^9$Li+n+n structure, where the shell-model description of $^9$Li involves all Slater determinants (90)
which can be built in the $p$ shell.
The oscillator parameter $b$
is taken identical for the three clusters. Going beyond this approximation raises enormous
difficulties due to center-of-mass problems, even in two-cluster calculations \cite{BK92}.
In Eq.\ (\ref{eq6}), the hypermoment $K$ runs from zero to infinity.  In practice a truncation value $K_{\rm max}$ is adopted.  
The hyperradial functions $\chi^{J\pi}_{\ell_x \ell_y K}(\rho)$ are to be determined from the Schr\"odinger equation (\ref{eq2}).

As for two-cluster systems, the RGM definition clearly displays the physical interpretation of the cluster approximation.  In practice, however, using the Generator Coordinate Method (GCM) wave functions is equivalent, and is more appropriate to systematic numerical calculations \cite{DD12}.  In the GCM, the wave function (\ref{eq6}) is equivalently written as  
\begin{align}
\Psi^{JM\pi}=\sum_{\gamma} \int dR \, f^{J\pi}_{\gamma}(R)
\, \Phi^{JM\pi}_{\gamma}(R),
\label{eq7}
\end{align}
where we use label $\gamma=(\ell_x, \ell_y, K)$. In this equation, $R$ is the generator coordinate, $\Phi^{JM\pi}_{\gamma}(R)$ are projected Slater determinants, 
and $f^{J\pi}_{\gamma}(R)$ are the generator functions (see Ref.\ \cite{DD12} for more detail).  
In practice, the integral is replaced by a sum over a finite set of $R$ values (typically 10-15 values are chosen, up to $R\approx 10-12$ fm).  

After discretization of (\ref{eq7}), the generator functions are obtained from the eigenvalue problem,
known as the Hill-Wheeler equation,
\begin{align}
	\sum_{\gamma n} & \biggl[
	H^{J\pi}_{\gamma,\gamma' }(R_n,R_{n'})
	- E^{J\pi}_{k} N^{J\pi}_{\gamma,\gamma' }(R_n,R_{n'})\biggr]
	f^{J\pi}_{(k) \gamma}(R_n)=0,
	\label{eq8}
\end{align}
where $k$ denotes the excitation level. The Hamiltonian and overlap kernels are obtained from 7-dimension 
integrals involving matrix elements between Slater determinants (see Refs.\ \cite{KD04,DD12} for detail).  
They are given by
\begin{align}
	N^{J\pi}_{\gamma,\gamma'}(R,R')&=
	\langle \Phi^{JM\pi}_{\gamma}(R) \vert \Phi^{JM\pi}_{\gamma' }(R') \rangle \nonumber \\
	H^{J\pi}_{\gamma,\gamma' }(R,R')&= 
	\langle \Phi^{JM\pi}_{\gamma}(R) \vert H \vert \Phi^{JM\pi}_{\gamma'}(R') \rangle.
	\label{eq9}
\end{align}
The non-projected matrix elements are computed with Brink's formula \cite{Br66}, and the main part of the 
numerical calculations is devoted to the two-body interaction which involves quadruple sums over the orbitals. 
The projection over angular momentum, which requires multidimension integrals, makes the calculation very 
demanding. A more detailed description is provided in Ref.\ \cite{De19}.

\subsection{Radii and energy curves}
From the wave function (\ref{eq6},\ref{eq7}), various properties can be computed. We discuss more
specifically the r.m.s. radius, defined as
\begin{align}
	\langle r^2\rangle=\frac{1}{A}\langle \Psi^{JM\pi} \vert \sum_{i=1}^A (r_i-R_{\rm c.m.})^2 \vert \Psi^{JM\pi} \rangle,
\label{eq10}
\end{align}
which, in the GCM is determined from
\begin{align}
&\langle r^2\rangle=\sum_{\gamma n}	\sum_{\gamma' n'} f^{J\pi}_{\gamma}(R_n) f^{J\pi}_{\gamma'}(R_n') \nonumber \\
&\hspace*{0.3 cm} \times \langle \Phi^{JM\pi}_{\gamma}(R_n) \vert \frac{1}{A} \sum_{i=1}^A (r_i-R_{\rm c.m.})^2 \vert
 \Phi^{JM\pi}_{\gamma'}(R_{n'}) \rangle.
\label{eq11}
\end{align}
The matrix elements between Slater determinants are obtained as in Eqs.\ (\ref{eq9}). Notice that these calculations
are rigorous for bound states, i.e. states with energy $E^{J\pi}_k$ lower than the three-cluster breakup threshold 
$E_T=E_1+E_2+E_3$, $E_i$ being the internal energy of cluster $i$, computed consistently with the same Hamiltonian. In this
case, the relative functions $\chi^{J\pi}_{\gamma}(\rho)$ [see Eq.\ (\ref{eq6})] tend rapidly to zero, and the sum over $(n,n')$ in
(\ref{eq11}) converges. For resonances ($E^{J\pi}_k > E_T$), the convergence of (\ref{eq11}) is not guaranteed. We discuss this
issue in more detail in Sect.\ \ref{sec3}.

The energies and r.m.s. radii discussed in the previous subsection involve all generator coordinates.
It is, however, useful to analyze the energies of the system for a single $R$ value. This quantity is
referred to as the energy curves. Two variants can be considered. In the former, only the
diagonal matrix elements of the Hamiltonian are used, and the energy curves are defined as
\begin{align}
E^{J\pi}_{\gamma}(R)=\frac{H^{J\pi}_{\gamma,\gamma}(R,R)}	{N^{J\pi}_{\gamma,\gamma}(R,R)}-E_T.
\label{eq12}
\end{align}
This definition ignores the couplings between the channels. At large distances, they tend to
\begin{align}
	E^{J\pi}_{\gamma}(R)\rightarrow \frac{Z_{\gamma \gamma}e^2}{R}+
	\frac{\hbar^2}{2m_N}\frac{(K+3/2)(K+5/2)}{R^2}+\frac{1}{4}\hbar \omega,
	\label{eq13}
\end{align}
where $Z_{\gamma \gamma}e^2$ is a diagonal element of the Coulomb three-body interaction
(see, for example, Ref.\ \cite{De10}), $m_N$ is the nucleon mass, and $\frac{1}{4}\hbar \omega$
is the residual energy associated with the harmonic oscillator functions ($\hbar \omega
=\hbar^2/m_N b^2$).

In the alternative approach, the energy curves stem from a diagonalization of the Hamiltonian
for a fixed $R$ value. They are given by the eigenvalue problem
\begin{align}
	\sum_{\gamma} & \biggl[
	H^{J\pi}_{\gamma ,\gamma'}(R,R)
	- E^{J\pi}_k(R) N^{J\pi}_{\gamma ,\gamma' }(R,R)\biggr] 
	c^{J\pi}_{(k) \gamma }(R)=0.
	\label{eq14}
\end{align}
At large distances, the coupling elements ($\gamma \neq \gamma'$) tend to zero and both definitions (\ref{eq12})
and (\ref{eq14}) are equivalent. In three-body systems, however, the couplings are known to extend
to large distances, even for short-range interactions (see, for example, Refs.\ 
\cite{DTB06,De10,DD09}).

The energy curves cannot be considered as genuine potentials. However, they provide various informations,
such as the existence of bound states or of narrow resonances, the level ordering, the cluster structure,
etc.

\section{Discussion of resonances}
\label{sec3}
The eigenvalue problem (\ref{eq8}) is, strictly speaking, valid for bound states only. The 
variational principle guarantees that an upper limit of the exact solution is found, and the wave function
tends exponentially to zero. The situation, however, is different for positive-energy states.
In that case, the lowest energy is zero, i.e. the optimal solution, according to the variational
principle, corresponds to a system where the clusters are at infinite distance from each other.

For narrow resonances, the bound-state approximation (BSA), whi\-ch is a direct extension
of (\ref{eq8}), usually provides a fair approximation of the energy, even with finite bases. If the
width is small, the energy is fairly stable when the basis changes. The situation is different for the
wave function. The long-range part may be sensitive to the choice of the basis, and matrix elements
using these wave functions may be unstable. A typical example will be shown with the $0^+_2$ resonance
of $\car$.

For broad resonances, there are various techniques which complement the variational method. The idea
is to avoid scattering calculations, such as in the $R$-matrix theory \cite{DB10}, where resonance properties
are derived from an analysis of the phase shifts (or scattering matrices). The complementary
methods have solid mathematical foundations, and are, in principle, relatively simple to
implement in variational calculations. We briefly summarize below some of them.

\begin{itemize}
\item The {\sl complex scaling method (CSM)} is based on the rotation of the space
and momentum coordinates \cite{Ho83,AC71,AMK06}. In other words, the space coordinate $\pmb{r}$ and momentum $\pmb{p}$ of each particle are transformed as
\begin{align}
	&U(\theta)\pmb{r}U^{-1}(\theta)=\pmb{r}\exp(i\theta), \nonumber \\
	&U(\theta)\pmb{p}U^{-1}(\theta)=\pmb{p}\exp(-i\theta),
	\label{eq15}
\end{align}
where $\theta$ is the rotation angle.  Under this transformation, the Schr\"{o}dinger equation reads
\begin{align}
	H(\theta)\Psi(\theta)=U(\theta)HU^{-1}(\theta)\Psi(\theta)=E(\theta)\Psi(\theta),
	\label{eq16}
\end{align}
and the solutions $\Psi(\theta)$ are square-integrable provided that $\theta$ is properly 
chosen \cite{AMK06}.  They can be expanded over a finite basis, after rotation of the Hamiltonian.  	Of course the potential should be available in an analytic form to apply transformation (\ref{eq15}).
	
The ABC theorem \cite{AC71} shows that the eigenvalues $E_{k}(\theta)$ are located on a straight 
line in the complex plane, rotated by an angle $2\theta$.  Resonant states are not affected by this angle 
and correspond to stable eigenvalues
\begin{align}
	E_k(\theta)=E_R-i\Gamma/2,
	\label{eq17}
\end{align}
where $E_R$ is the energy and $\Gamma$ the width of the resonance.
	
Recently, the CSM has been extended to the calculation of level densities \cite{SMK05,SKG08} and 
to dipole strength distributions \cite{AMK06}.  Of course the resonance properties 
(\ref{eq17}) derived from the CSM should also be consistent with those derived from a phase-shift analysis.\\

\item In the {\sl complex absorbing potential (CAP)} method, an imaginary potential is added to the Hamiltonian kernel. The first applications were developed in atomic physics \cite{RM93} and in
non-microscopic nuclear models \cite{TOI14}. Ito and Yabana \cite{IY05} have extended the method
to microscopic cluster calculations within the GCM. The Hamiltonian kernel (\ref{eq9}) is replaced by
\begin{align}
	H(R,R')\rightarrow H(R,R')-i\eta W(R)\delta(R-R'),
	\label{eq18}
\end{align}
where $\eta$ is a positive real number. The absorbing potential is usually taken as
\begin{align}
	W(R)=\theta(R-R_0)(R-R_0)^{\beta},
	\label{eq19}
\end{align}
where $R_0$ is an arbitrary radius, larger than the range of the nuclear force and $\theta$ is the step
function. In most calculations, $\beta$ is taken as $\beta=4$. 

This method provides the energy and widths of resonances as in (\ref{eq17}), even for broad states. In Ref.\ \cite{IY05}, it was shown,
however, that the method needs many generator coordinates. In their microscopic study of
$\alpha+^6$He scattering, Ito and Yabana use 100 generator coordinates, up to $R=50$ fm. For
computational reasons, this method is difficult to apply to three-body systems, owing to the strong couplings between the channels, and to the long range of the potentials.\\

\item The {\sl analytic continuation in a coupling constant (ACCC)} method has been proposed by Kukulin {\sl et al.}
\cite{KKH89} to evaluate the energy and width of a resonance. The main advantage is that the ACCC method only requires bound-state calculations, much simpler than scattering calculations involving boundary conditions. Some applications to a microscopic description of two- and three-cluster models can be found in Ref.~\cite{TSV99}. The
method has been applied to a non-microscopic description of $\car$ in Ref.\ \cite{KK05}.

To apply the ACCC method, one assumes that the Hamiltonian can be written as
\begin{align}
H(u)=H_0+u\, H_1,
\label{eq20}
\end{align}
where $u$ is a linear parameter. The linear part $H_1$ is supposed to be attractive so that, for increasing $u$ values, the system becomes bound. For $u=u_0$, the energy is zero, and we have
$E(u_0)=0$.

The problem is to determine the resonance properties for the physical value $u<u_0$. In the bound-state regime $(u>u_0)$, the wave number $k$ is imaginary, and is parametrized by a Pad\'e approximant as
\begin{align}
k(x)=i\frac{c_0+c_1x+\cdots+c_M x^M}{1+d_1x+\cdots+d_N x^N},
\label{eq22}
\end{align}
where $x=\sqrt{u-u_0}$, and ($M,N$) define the degree of the Pad\'e approximant. The ($M+N+1$) coefficients $c_i$ and $d_j$ are calculated in the bound-state region by using $u_i$ values such that $E(u_i)<0$. Going to the physical $u$ value ($u<u_0,x$ imaginary), one determines $k$ from (\ref{eq22}). The energy $E_R$ and width $\Gamma$ are obtained from
\begin{align}
E=\frac{\hbar^2 k^2}{2m}=E_R-i\Gamma /2.
\label{eq23}
\end{align}
The ACCC is, in principle, a simple extension of the variational method. 
However, it was pointed out \cite{KKH89,TSV99} that this method requires a high accuracy in the numerical calculation. In particular, the $u_0$ value must be determined with several digits. It is therefore not realistic for microscopic three-body systems.\\

\item The {\sl box method}  \cite{MCD80} can be used for narrow resonances.  This method has been 
applied essentially in atomic physics.  The idea is to search for positive 
eigenvalues of the Schr\"odinger equation (\ref{eq2}) inside a box.  
Then, looking at the eigenvalues as a function of the box size, a narrow resonance
appears at a stable energy (see, for example, Fig. 1 of Ref.\ \cite{MCD80}).  
This method is simpler than other approximate techniques and, therefore, 
permits the use of large bases. The method has been recently extended to the determination of resonance
widths \cite{ZMZ09}. However, it requires the numerical calculation of the first and second derivatives, which 
means that, in practice, many generator coordinates must be used for a good accuracy of the resonance parameters.

\end{itemize}

\section{Application to $\car$}
\label{sec4}
The $\car$ nucleus has attracted much attention in recent years, in particular for the $0^+_2$ resonance,
known as the Hoyle state (see Ref.\ \cite{FF14} for a recent review). The Hoyle state,
located just above the $3\alpha$ threshold ($E_R=0.36$ MeV), is quite important in nuclear
astrophysics since its properties determine the triple-$\alpha$ reaction rate. Its existence
was predicted by Hoyle \cite{Ho54} on the basis of the observed abundance of $\car$.

There is an impressive literature about the Hoyle state, and we refer to Refs.\ \cite{FF14,ZFH20} for an
overview. One of its characteristics is to present a marked $\alpha+^8$Be cluster structure \cite{DB87b}.
This $\alpha$ clustering is well established in many light nuclei, such as $^7$Li, $^{7,8,9,10}$Be,
$^{16,17,18}$O, $^{18,19,20}$Ne \cite{FHI80}. The specificity of the $\car$ nucleus is that the
second cluster $^8$Be also presents an $\alpha$ cluster structure. As in all excited states
located near a breakup threshold, the Hoyle state presents an extended density, which means that
the density at short distances is decreased if compared to well bound states. This natural property,
common to all nuclei, lead some authors to refer to the concept of "dilute gas" \cite{Ka07,Fu15} and of
Bose-Einstein Condensates \cite{THS01,ZFH20}.

Our aim here is not to perform new calculations on the $\car$ nucleus. The first microscopic
$3\alpha$ calculation was performed by Uegaki {\sl et al.} in 1977 \cite{UOA77}, and improved
in different ways \cite{Ka81,DB87b,THS01,CFN07,Ka07,SMO08}. The {\sl ab initio} calculation of
Ref.\ \cite{CFN07} works rather well for the ground state, but needs the introduction of 
specific $3\alpha$ configurations to reproduce the Hoyle state. One of the frequent issues about the Hoyle state is the determination of its r.m.s. radius, which is expected to be large
(see references in Ref.\ \cite{FF14}).

We adopt the same microscopic $3\alpha$ model as in Ref.\ \cite{SMO08}, where the Minnesota
nucleon-nucleon interaction \cite{TLT77} with $u=0.9487$ was used. For the generator coordinates, we 
use a large basis: 12 values from 1.5 to 18 fm with a step of 1.5 fm, and 3 additional values at 20, 22
and 24 fm. This basis is unusually large, but permits a reliable discussion concerning the properties
of resonances. 

\begin{figure}[htb]
	\begin{center}
		\epsfig{file=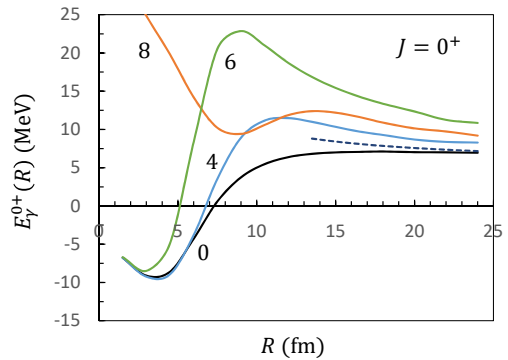,width=6.5cm}
		\caption{Energy curves (\ref{eq12}) for $J=0^+$ in $\car$. The curves are labeled by the
			$K$ values. The dashed line	represents the asymptotic behaviour (\ref{eq13}) for $K=0$.}
		\label{fig_c12_ce}
	\end{center}
\end{figure}

In Fig.\ \ref{fig_c12_ce}, we display the energy curves (\ref{eq12}) for $J=0^+$ and for different
$K$ values. Note that in $\car$, a full symmetrization of the wave functions for $\alpha$ exchange
leads to the cancellation of the $K=2$ component (see, for example, Ref.\ \cite{De10}). The dashed line
illustrates the asymptotic behaviour (\ref{eq13}) with $Z_{00}=28.81$ \cite{De10}. The energy curves for
$K=0,4,6$ present a minimum around $R=4$ fm. At short distances, the antisymmetrization between the nucleons makes these curves equivalent. This is not true at large distances, where the effective
three-body Coulomb interaction is different, and where the centrifugal term plays a role.

\begin{figure}[htb]
	\begin{center}
		\epsfig{file=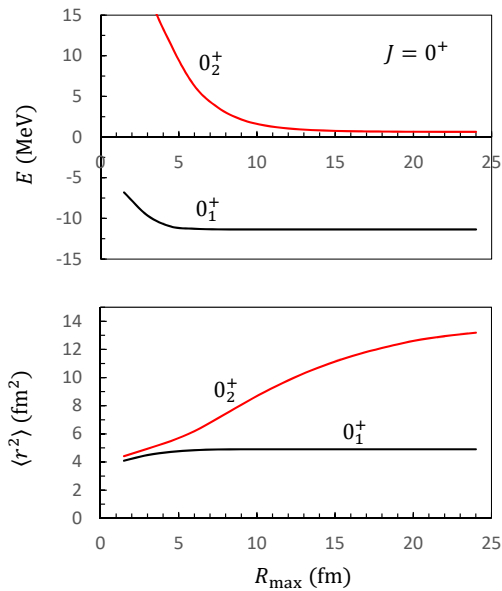,width=6.5cm}
		\caption{Total energy (top) and r.m.s. radius (bottom) of the $0^+_1$ and $0^+_2$
			states in $\car$. Energies are given with respect to the $3\alpha$ threshold.}
		\label{fig_c12_0}
	\end{center}
\end{figure}

Figure \ref{fig_c12_0} presents the binding energy (with respect to the $3\alpha$ threshold) of
the $0^+_1$ and $0^+_2$ states for increasing size of the basis. We define $\rmax$ as the maximum $R$
value included in the basis. The upper and lower panels present the energy and r.m.s. radius, 
respectively. As expected for a bound state, the $0^+_1$ energy and r.m.s. radius converge rapidly. With $\rmax \approx 6$ fm, a fair convergence is reached. The corresponding radius $\sqrt{<r^2>}=2.21$ fm is too
small compared to experiment (2.48 fm \cite{KPS17}) but this difference is due to the overbinding of the
theoretical state.

More interesting results are obtained for the $0^+_2$ state. If the energy remains almost constant
above $\rmax \approx 15$ fm, the r.m.s. radius significantly increases with larger bases. We go here
up to $\rmax=24$ fm, which represents a huge value, much larger than in standard calculations
(see, for example, Ref.\ \cite{SMO08}). This is, however, necessary to illustrate convergence issues.
As we may expect from the unbound nature of the $0^+_2$ state, the r.m.s. radius diverges. Any (large)
value can be obtained, provided that the basis is large enough. This explains why calculations
in the literature are quite different \cite{FF14}.
 
\begin{figure}[htb]
	\begin{center}
		\epsfig{file=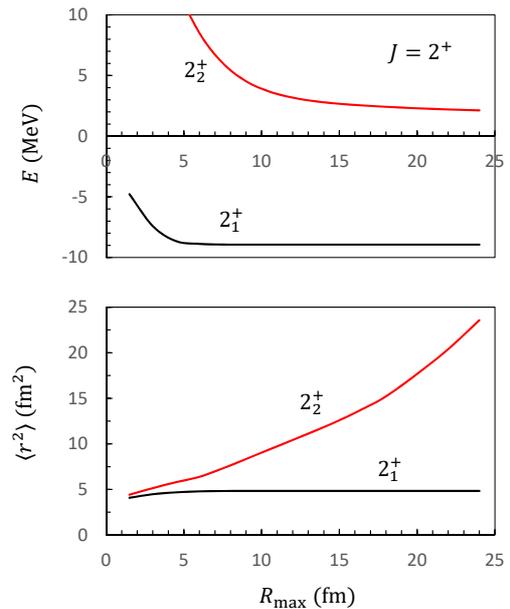,width=6.5cm}
		\caption{See caption to Fig.\ \ref{fig_c12_0} for $J=2^+$.}
		\label{fig_c12_2}
	\end{center}
\end{figure}

The same quantities are plotted in Fig.\ \ref{fig_c12_2} for $J=2^+$. The existence of a $2^+_2$ resonance, second state of a band based on the Hoyle state, is well established \cite{Ka81,DB87b}, but this state
should be broad. Even the energy is not stable, and does not present a plateau with $\rmax$. The
corresponding r.m.s. radius strongly diverges. Notice that a microscopic $3\alpha$ model
does not predict any $0^+_3$ or $2^+_3$ resonance.

Let us briefly comment on non-microscopic descriptions of the $3\alpha$ system. Two $\alpha+\alpha$ potentials,
the shallow Ali-Bodmer potential \cite{AB66} and the deep Buck potential \cite{BFW77} reproduce very well
the experimental $\alpha+\alpha$ phase shifts up to 20 MeV. When applied to $N\alpha$ systems ($N>2$), however,
none of these potentials provides satisfactory results. With the Ali-Bodmer potential, the $\car$ ground state
is strongly underbound \cite{TBD03}, and the $0^+_2$ resonance is far above the $3\alpha$ threshold, in
contradiction with experiment. This problem may be partly addressed by adding a phenomenological 3-body force, but
this technique introduces spurious $3\alpha$ resonances \cite{De10}.

The deep Buck potential raises the question of the two-body forbidden states which should be removed in the
$3\alpha$ model. This is usually done by using a projection technique but, here also, several problems
remain (see the discussions in Refs.\ \cite{FMK04,De10}). An efficient alternative would be to use non-local 
$\alpha+\alpha$ potentials \cite{SMO08}, but the simplicity of non-microscopic models is lost. A fully
satisfactory description of the $3\alpha$ system within non-microscopic models remains an open issue.

\section{Application to $\mg$}
\label{sec5}
The $\oaa$ system is more complex than $\car$ since the
level density is much higher. The $\mg$ nucleus was studied within an $\ane$ multicluster model in Ref.\  \cite{DB89b}, 
where $^{20}$Ne is described by
an $\alpha+^{16}$O structure. In Ref.\ \cite{IIK11} the authors use a microscopic $\oaa$
model to search for $0^+$ resonances in a stochastic approach. The basis, however, is more limited
since a fixed geometry is adopted. The present work contains a more extended basis, owing to the use
of the hyperspherical coordinates. 

The oscillator parameter is chosen as $b=1.65$ fm, which
represents a compromise between the optimal values of $\alpha$ and of $^{16}$O. We use the Volkov force
V2 \cite{Vo65} with a Majorana exchange parameter $M=0.624$. This value reproduces the binding
energy of the ground state with respect to the $\oaa$ threshold ($-14.05$ MeV). 

The calculations of the matrix elements are much longer than for $\car$. Since most of the computer
time is devoted to the quadruple sums involved in the trwo-body interaction, the ratio
between the computer times is approximately given by $6^4/3^4=16$, since $\mg$ involves 6
orbitals ($s$ and $p$), whereas $\car$ involves 3 orbitals ($s$ only).

\begin{figure}[htb]
	\begin{center}
		\epsfig{file=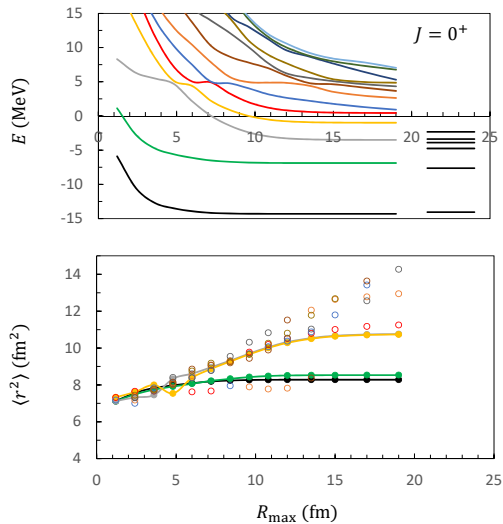,width=6.5cm}
		\caption{Energies with respect to the $\oaa$ threshold (top) and r.m.s. radii (bottom)
			of $J=0^+$ states in $\mg$ for different sizes of the basis. The color schemes are identical in both panels. In the upper panel,
			the right side shows experimental energies. In the lower panel, the r.m.s. radii for the third and fourth eigenvalues
		are almost superimposed.}
		\label{fig_mg24_0}
	\end{center}
\end{figure}

As for $\car$, we use a large basis with 10 $R$-values from 1.2 fm to 12 fm, complemented by larger
values $R=13.5,15,17,19$ fm. In Fig.\ \ref{fig_mg24_0}, we present the binding energies and r.m.s. radii 
of $J=0^+$ states as a function of $\rmax$, the maximum $R$ value included in the basis.

The upper panel shows that four states are bound with respect to the $\oaa$ decay, two of which are
above the $\ane$ threshold ($-4.7$ MeV), and are therefore not rigorously bound. This is
confirmed by the r.m.s. radii (lower panel). The two lowest states converge rapidly to
$<r^2>\approx 8$ fm$^2$, whereas the $0^+_3$ and $0^+_4$ radii present a slower convergence
and a larger radius $<r^2>\approx 10$ fm$^2$. The radii corresponding to the four negative-energy curves are linked by solid lines. 

The right side of the upper panel shows the experimental
$0^+$ states. The ground-state energy is adjusted by the nucleon-nucleon interaction. The $0^+_2$
energy is in fair agreement with experiment. In the high-energy part of the spectrum, several
$0^+$ experimental states are present.

Figure \ref{fig_mg24_0} suggests two important properties. Around $E\approx 5$ MeV, there is
a plateau in the energy curves, and the corresponding radii are rather insensitive to the size
of the basis. This is consistent with a high-energy ($E_x\approx 19$ MeV) resonance in the
$\mg$ spectrum. As $\rmax$ increases, the label of the eigenvalue varies. The radii corresponding to the plateau therefore show up as individual points around $<r^2>\approx 7.5$ fm$^2$. The dip near $\rmax=5$ fm is due to a crossing between the energy curves.
The second information concerns the radii. The lower panel of Fig.\ \ref{fig_mg24_0}
shows a clear distinction between physical states and approximations of the continuum. The former
present a stable radius, whereas the latter are characterized by diverging radii. A careful study
of the radii, and in particular of their stability against the extension of the basis, is therefore
an efficient way to make a distinction between physical states and pseudostates.

\begin{figure}[htb]
	\begin{center}
		\epsfig{file=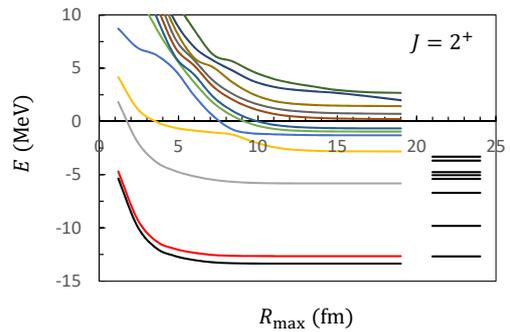,width=6.5cm}
		\caption{See caption to Fig.\ \ref{fig_mg24_0} for $J=2^+$. Only energies are displayed.}
		\label{fig_mg24_2}
	\end{center}
\end{figure}

Figure \ref{fig_mg24_2} displays the energy curves for $J=2^+$. The level density is still
higher than for $J=0^+$ and it is difficult to make a clear link between theory and experiment.
The $2^+_1$ and $2^+_3$ energies are in fair agreement, but the GCM $2^+_2$ energy is too low by about 2 MeV. This is probably due to the lack of a spin-orbit force in a $\oaa$ model. The model predicts
seven states below the $\oaa$ threshold. The r.m.s. radii are not presented as they qualitatively follow those
of $J=0^+$. The converged radii for the $2^+_1$ and $2^+_2$ states are around $<r^2>\approx 8.2$ fm$^2$, which
is close to the radius of the ground state.

\section{Conclusion}
\label{sec6}
The goal of the present work is to illustrate the calculation of resonance properties in cluster models,
and more especially in multicluster models. A rigorous treatment of the resonances would require a
scattering theory, with exact boundary conditions. If this approach is fairly simple in two-cluster
models, it raises strong difficulties when more than two clusters are involved. The bound-state
approximation is commonly used owing to its simplicity. We have shown, however, that positive-energy eigenvalues
should be treated carefully, and that, even for narrow resonances, the wave function may be sensitive to the basis.
A direct consequence is that some properties, such as the r.m.s. radius, are unstable. The stability against the
basis should be assessed.

Several method exist to complement the bound-state approximation, such as the CSM or the ACCC. They permit to
determine the energy and width of a resonance. In practice, however, they are difficult to apply to
microscopic calculations since they usually need very large bases. We have used an alternative of the
box method, where the number of basis functions is progressively increased. We have shown that a stability
of the energy can be obtained, but the corresponding r.m.s. radii are unstable. The application to the
Hoyle state in $\car$ is an excellent example which explains the variety of the values in the literature.

\section*{Acknowledgments}
This work was supported by the Fonds de la Recherche Scientifique - FNRS under Grant Numbers 4.45.10.08 and J.0049.19.
It benefited from computational resources made available on the Tier-1 supercomputer of the 
F\'ed\'eration Wallonie-Bruxelles, infrastructure funded by the Walloon Region under the grant agreement No. 1117545. 

%

\end{document}